\documentclass[8.5pt,twoside,twocolumn]{article}
\usepackage[super,sort&compress,comma]{natbib} 
\usepackage{mhchem}
\usepackage{times,mathptmx}
\usepackage{sectsty}
\usepackage{balance} 

\usepackage{graphicx} 
\usepackage{lastpage}
\usepackage[format=plain,justification=raggedright,singlelinecheck=false,font=small,labelfont=bf,labelsep=space]{caption} 
\usepackage{fancyhdr}
\usepackage{amsmath,amssymb}
\usepackage{amsmath,amssymb}
\newcommand{\fig}[1]{fig~\ref{#1}}
\newcommand{\figa}[1]{fig~\ref{#1}a}
\newcommand{\figb}[1]{fig~\ref{#1}b}
\newcommand{\figc}[1]{fig~\ref{#1}c}

\newcommand{\Figa}[1]{Fig~\ref{#1}a}

\newcommand{\Figc}[1]{Fig~\ref{#1}c}

\newcommand{\sct}[1]{section~\ref{#1}}

\begin{document}

\thispagestyle{plain}
\fancypagestyle{plain}{
\fancyhead[L]{\includegraphics[height=8pt]{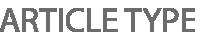}}
\fancyhead[C]{\hspace{-1cm}\includegraphics[height=20pt]{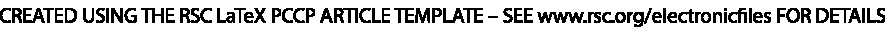}}
\fancyhead[R]{\includegraphics[height=10pt]{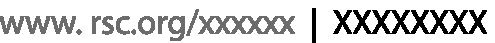}\vspace{-0.2cm}}
\renewcommand{\headrulewidth}{1pt}}
\renewcommand{\thefootnote}{\fnsymbol{footnote}}
\renewcommand\footnoterule{\vspace*{1pt}%
\hrule width 3.4in height 0.4pt \vspace*{5pt}} 
\setcounter{secnumdepth}{5}

\makeatletter 
\def\subsubsection{\@startsection{subsubsection}{3}{10pt}{-1.25ex plus -1ex minus -.1ex}{0ex plus 0ex}{\normalsize\bf}} 
\def\paragraph{\@startsection{paragraph}{4}{10pt}{-1.25ex plus -1ex minus -.1ex}{0ex plus 0ex}{\normalsize\textit}} 
\renewcommand\@biblabel[1]{#1}            
\renewcommand\@makefntext[1]%
{\noindent\makebox[0pt][r]{\@thefnmark\,}#1}
\makeatother 
\renewcommand{\figurename}{\small{Fig.}~}
\sectionfont{\large}
\subsectionfont{\normalsize} 

\fancyfoot{}
\fancyfoot[LO,RE]{\vspace{-7pt}\includegraphics[height=9pt]{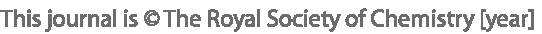}}
\fancyfoot[CO]{\vspace{-7.2pt}\hspace{12.2cm}\includegraphics{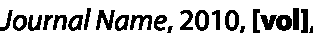}}
\fancyfoot[CE]{\vspace{-7.5pt}\hspace{-13.5cm}\includegraphics{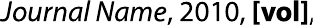}}
\fancyfoot[RO]{\footnotesize{\sffamily{1--\pageref{LastPage} ~\textbar  \hspace{2pt}\thepage}}}
\fancyfoot[LE]{\footnotesize{\sffamily{\thepage~\textbar\hspace{3.45cm} 1--\pageref{LastPage}}}}
\fancyhead{}
\renewcommand{\headrulewidth}{1pt} 
\renewcommand{\footrulewidth}{1pt}
\setlength{\arrayrulewidth}{1pt}
\setlength{\columnsep}{6.5mm}
\setlength\bibsep{1pt}

\twocolumn[
  \begin{@twocolumnfalse}
\noindent\LARGE{\textbf{Viscous fingering at ultralow interfacial tension$^\dag$}}
\vspace{0.6cm}

\noindent\large{\textbf{Siti Aminah Setu,\textit{$^{a,\dag}$} Ioannis Zacharoudiou,\textit{$^{b,\dag}$} Gareth J. Davies,\textit{$^{a}$} Denis Bartolo,\textit{$^{c}$} S\'ebastien Moulinet,\textit{$^{d}$}  Ard A. Louis,\textit{$^{b}$} Julia M. Yeomans,\textit{$^{b}$}, and Dirk G.A.L. Aarts$^{\ast}$\textit{$^{a}$}}}\vspace{0.5cm}

\noindent\textit{\small{\textbf{Received Xth XXXXXXXXXX 20XX, Accepted Xth XXXXXXXXX 20XX\newline
First published on the web Xth XXXXXXXXXX 200X}}}

\noindent \textbf{\small{DOI: 10.1039/b000000x}}
\vspace{0.6cm}

\noindent \normalsize{We experimentally study the viscous fingering instability in a fluid-fluid phase separated colloid-polymer mixture by means of laser scanning confocal microscopy and microfluidics. We focus on three aspects of the instability. (i) The interface between the two demixed phases has an ultralow surface tension, such that we can address the role of thermal interface fluctuations. (ii) We image the interface in three dimensions allowing us to study the interplay between interface curvature and flow. (iii) The displacing fluid wets all walls completely, in contrast to traditional viscous fingering experiments, in which the displaced fluid wets the walls. We also perform lattice Boltzmann simulations, which help to interpret the experimental observations. 
}
\vspace{0.5cm}
 \end{@twocolumnfalse}
  ]

\section{Introduction}
\footnotetext{\dag~Electronic Supplementary Information (ESI) available: [details of any supplementary information available should be included here]. See DOI: 10.1039/b000000x/}


\footnotetext{\textit{$^{a}$~Department of Chemistry, Physical and Theoretical Chemistry Laboratory, University of Oxford, South Parks Road, Oxford OX1 3QZ, UK E-mail: dirk.aarts@chem.ox.ac.uk}}
\footnotetext{\textit{$^{b}$~Rudolf Peierls Centre for Theoretical Physics, University of Oxford, 1 Keble Road, OX1 3NP, Oxford , UK. }}
\footnotetext{\textit{$^{c}$~Laboratoire de physique et m\'ecanique des milieux h\'et\'erog\`ene, PMMH ESPCI-CNRS UMR 7636-P6-P7, Paris, France }}
\footnotetext{\textit{$^{d}$~Laboratoire de Physique Statistique, Ecole Normale Sup\'erieure, 24, Rue Lhomond, F-75231 Paris Cedex 05, France }}


\footnotetext{\dag~These authors contributed equally to this work}

Whenever a low viscosity fluid displaces a high viscosity fluid in a porous medium or Hele-Shaw cell, the interface is rendered unstable and viscous fingers develop, grow and compete. The first detailed experimental and theoretical account of this instability was given by Saffman and Taylor in their seminal 1958 paper\cite{Saffman_Taylor_1958}. Since then the viscous fingering instability has witnessed intense study culminating in breakthroughs in theory and in experiment in the 1980s dealing especially with the role of surface tension\cite{mclean81,Tabeling_Libchaber1986, shraiman86, hong86, PhysRevLett.56.2036,pelce88,couder00}. Here, it was realised that the surface tension acts as a singular perturbation in the Saffman-Taylor problem and specifically in its McLean-Saffman solutions, see for example \cite{bensimon86,homsy87} for reviews of the problems. Recently, viscous fingering has witnessed a renewed interest, for example in the role of visco-elastic effects \cite{lindner09}, the importance of three dimensional phenomena \cite{ledesma_aguilar_pagonabarraga_2}, possible stabilization mechanisms through channel geometries \cite{stone12}, and as a means of providing efficient mixing in microfluidics channels \cite{jha11}. The classic zero-surface-tension limit also continues to inspire experimental studies, for example on granular matter \cite{cheng08} where instead of rounded fingers a fractal structure with sharp cusps develops. 

\begin{figure}
\includegraphics[width=80mm]{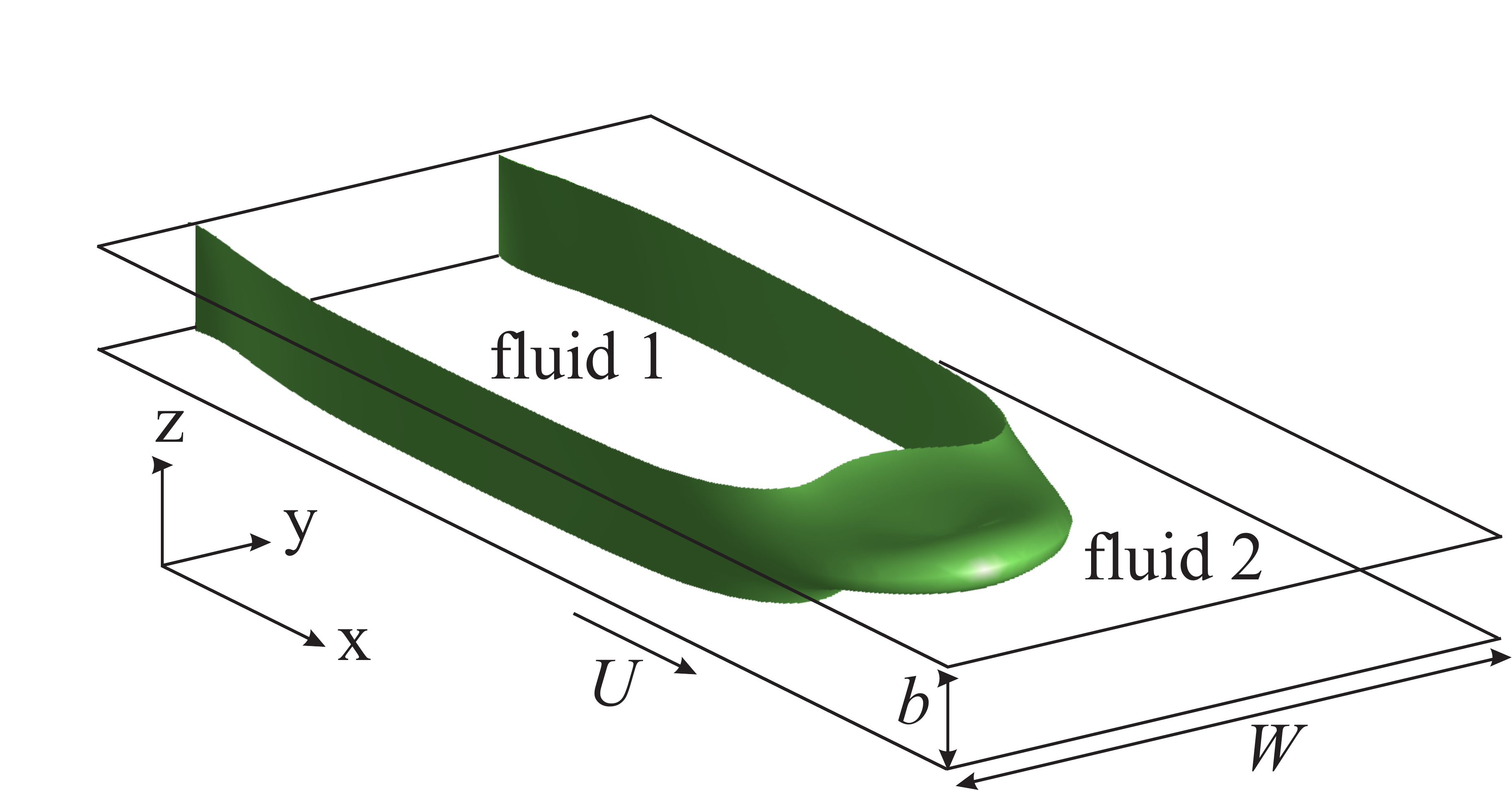}
\caption{Schematic depiction of a computer-generated viscous finger in a Hele-Shaw geometry, with fluid 1 (low viscosity) displacing fluid 2 (high viscosity). Microscopy observations will be made in the $xy$-plane at $z=b/2$ and in the $xz$-plane through the tip of the finger. }
\label{figsketch}
\end{figure}

Experiments on the viscous fingering instability are typically performed using Hele-Shaw cells with an aspect ratio $\epsilon = b/W \ll 1$, with $b$ the channel thickness and $W$ the channel width. In this limit, by averaging over a flow profile which is assumed to be parabolic across the narrow direction $z$ (perpendicular to the $xy$-plane of the viscous fingers, see \fig{figsketch}), the problem is effectively rendered two dimensional and McLean \& Saffman\cite{mclean81} showed that there is a unique solution for the relative finger width $\lambda$ (with respect to the channel width), which  is governed by a dimensionless parameter \cite{tryggvason83} $1/B = 12 Ca_{xy}/\epsilon^2$. Here, the capillary number $Ca_{xy}$ is defined as $Ca_{xy} = (\eta_2 -  \eta_1 ) U / \gamma $, with $U$ the velocity of the tip of the finger, $\gamma$ the surface or interfacial tension, and $\eta_1$ and $\eta_2$ the viscosities of the displacing  and displaced fluids, respectively, where $\eta_1 < \eta_2$. 

The numerical results of McLean and Saffman suggested that the relative finger width $\lambda$ is a monotonically decreasing function of $1/B$ that tends to an asymptotic value of 1/2. Experiments carried out by Tabeling and Libchaber showed that introducing this control parameter alone is not enough to fully describe the instability\cite{Tabeling_Libchaber1986}. They showed that this is due to a thin layer of the more viscous fluid, adhered to the walls and left behind the advancing finger, which changes the curvature of the interface and renders the problem three dimensional. Using the theoretical prediction for the pressure drop across the interface\cite{Park_Homsy_1984} to rescale the surface tension indeed gave much better agreement with the predictions for low $1/B$.  
However, for larger $1/B$ the experiments \cite{Tabeling_Zocchi_Libchaber_1987} showed that the relative finger width can fall below the $\lambda=1/2$  limit predicted by both the original and modified two dimensional theories\cite{mclean81}.

In this paper we study the viscous fingering instability by using colloidal fluids, which enables us to revisit some of the problems introduced above and in particular address the following three different issues: (i) The interface of the phase separated colloidal fluids is characterized by an ultralow interfacial tension, which makes it possible to directly observe thermal capillary waves at the interface \cite{capwaves}. Thermal interface fluctuations may modify continuum hydrodynamics \cite{moseler00,eggers02,grun06} and hence the form of the instability; moreover, the experiments may shine light on the zero-surface-tension limit mentioned above. (ii) With laser scanning confocal microscopy we are able to observe the finger in all three dimensions, including the curvature of the interface in the $xz$-plane. (iii) In our system the low viscosity, displacing phase, completely wets the confining walls, which is the opposite wetting situation to traditional Saffman-Taylor experiments, where the displaced fluid wets the walls and leaves behind a thin film as explained above. This directly implies that the interface curvature in the $xz$-plane will play a different role than in the experiments of Tabeling and Libchaber \cite{Tabeling_Libchaber1986}. We furthermore note that our wetting situation is similar to imbibition experiments \cite{alava04}, where a liquid is sucked up in e.g. a porous medium by capillary effects; however, we push the displacing fluid, which has a lower viscosity than the displaced fluid, at velocities at least an order of magnitude above the spontaneous imbibition velocity and the observed phenomena are therefore best analyzed within the viscous fingering framework. 

To further elucidate our experimental findings we will compare experiments to lattice Boltzmann simulations, which effectively solve the Navier-Stokes equations and provide detailed information about the various physical parameters at play, which can be varied over a wide range.

The paper is organized as follows; in the next section we will provide the experimental details. In \sct{resul} we will present and discuss the results. In \sct{concl} conclusions will be drawn.

\section{Experimental details}

As our model fluid we used a mixture of colloids and non-adsorbing polymers, which separated into a colloidal liquid phase (colloid-rich, polymer-poor) in coexistence with a colloidal gas phase (colloid-poor, polymer-rich) due to the depletion interaction\cite{asakura:1255, Asakura1958, Vrij}. We used dispersions of fluorescently labelled PMMA particles with a diameter of 210 nm in water. A xanthan polymer solution ($M_{w} = 4 $x$ 10^{6}$ g mol$^{-1}$, $R_{g}=264 $ nm) was added to the dispersion to create a phase separating mixture. The phase diagram and experimental details have been published elsewhere\cite{Jamie_Howe_Davies_Dullens_Aarts}. This provides an excellent model system to study properties of fluid-fluid interfaces that are not accessible at the molecular scale\cite{DirkAarts05072004}. Because we are dealing with a colloidal, rather than a molecular fluid, the interface width is $\sim \mu$m, rather than a few nm and surface tensions are reduced by a factor $\sim 10^6$. The surface tensions $\gamma$ were obtained from the capillary wave spectrum\cite{DirkAarts05072004}. We will focus on a statepoint for which $\gamma$=30 nN/m. A rheological characterization was performed on a TA AR-G2 rheometer giving a viscosity contrast ratio $r = (\eta_2-\eta_1)/(\eta_2+\eta_1) = 0.4$. Noticeable shear-thinning was observed for the colloidal gas phase from a shear-rate of 10 s$^{-1}$, which was above the maximum shear rate in the experiment (estimated as $U/b \sim 0.3 s^{-1}$). 

\begin{figure}
\includegraphics[width=75mm]{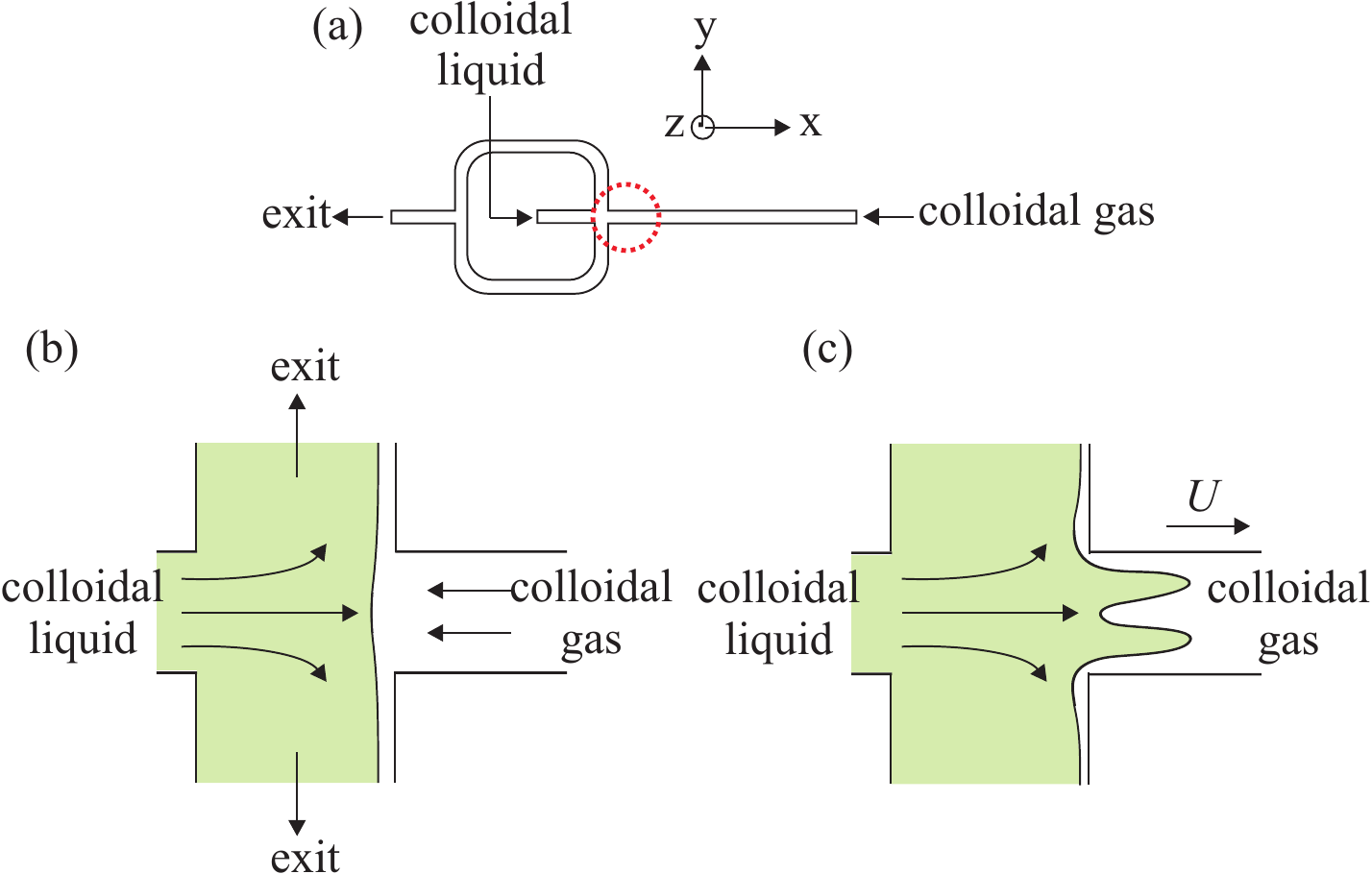}
\caption{(a) Schematic depiction of the microfluidic set-up, which enables us to start with a flat interface that is subsequently pushed into a long, flat channel, where the viscous fingers develop. The bottom two panels represent the zoomed in area indicated by the (red/dashed) circle, with (b) setting up the interface and (c) initiating fingering. }
\label{figsetup}
\end{figure}

We brought the separated colloidal fluids together in a microfluidic device, sketched in \fig{figsetup}, which was designed such that a straight colloidal interface could be formed, before it entered the Hele-Shaw geometry. We used channel thicknesses of 10, 14 and 17 $\mu$m. The capillary length in our system was 14 $\mu$m, such that gravitational effects in the channel played only a marginal role; for a study of the behaviour of a meniscus between two parallel plates, see Ref. \cite{jamieprl}. The width of the channel was varied between 100 $\mu$m and 1 mm. Gravity was used to produce pressure-driven flows. In this experimental setup the interface velocity slowly increased, but at any given time the finger was in a quasi-stationary state. The velocity $U$ was measured directly from consecutive confocal microscopy images. We used laser scanning confocal microscopy (LSCM; Zeiss Exciter), which recorded the fluorescence of excited dye within the colloids such that the colloid-rich phase will be bright and the polymer-rich phase dark, to image the flow of the colloidal phases in three dimensions. The colloid-rich phase, which had a lower viscosity $\eta_1$, was pushed into the long channel containing the polymer-rich phase (with viscosity $\eta_2$), from which point the development of viscous fingers was recorded. 

\section{Results and discussion}
\label{resul}
In this section we will present and discuss the results. We will start with (a) fingering in the $xy$-plane, the wide dimension of the channel. This will be followed by (b) the observations in the $xz$-plane, where we `make a cut' through the finger as it develops. We will focus on the interfacial curvature along $z$. Finally, we will discuss (c) the interplay between the curvature in the $z$-direction and the fingering in the $xy$-plane. 

\begin{figure}
\includegraphics[width=75mm]{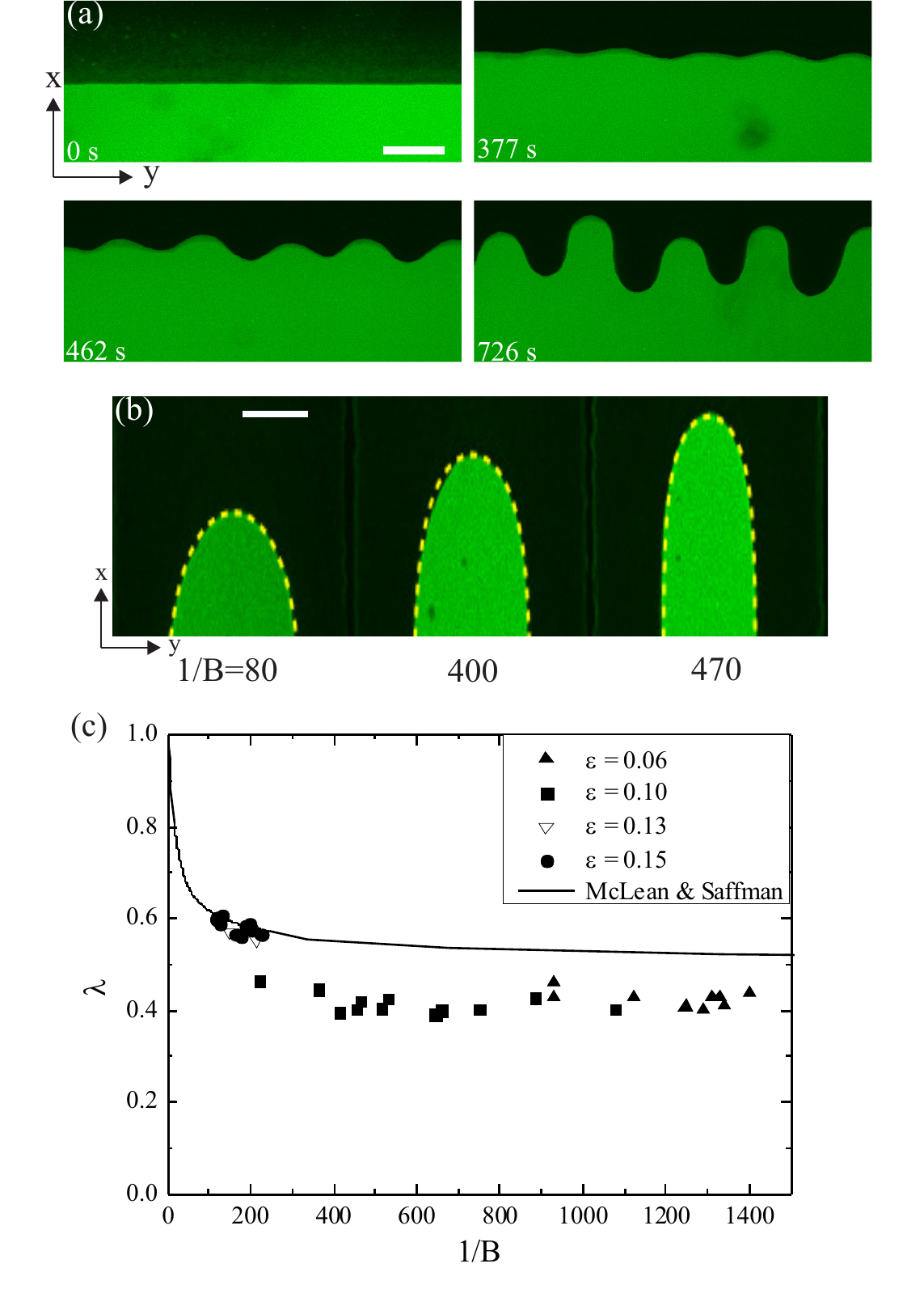}
\caption{Viscous fingering in the $xy$-plane. (a) LSCM images of the viscous fingering instability in a wide channel, where many fingers develop. (b) LSCM images of a single finger at different modified capillary numbers $1/B$. The dashed curve follows from the Pitts equation\cite{Pitts_1980}.  (c) Scaled finger width $\lambda$ as a function of $1/B$ for various aspect ratios. The full curve represents the McLean-Saffman solution \cite{mclean81}. Scale bars indicate 50 $\mu$m.}
\label{figxyfinger}
\end{figure}

{\bf (a) Fingering in the $xy$-plane:}
\Figa{figxyfinger} shows laser scanning confocal microscopy images of the viscous-fingering instability  in a wide channel ($W = 1$ mm) that can accommodate many fingers. The fingering has all the characteristics of the Saffman-Taylor instability and the further development of the fingers, where they interact and show tip-splitting and side-braching effects, is very rich, but lies outside the scope of the current work. Here we focus on the fingering in smaller channels between $W = 100~\mu$m up to 160 $\mu$m with channel thicknesses $b =$ 10, 14 or 17 $~\mu$m, where only a single finger is present, see \figb{figxyfinger}. 
Despite the ultralow value of the surface tension $\gamma$, it does act as a singular perturbation on the instability and a single finger width $\lambda$ (measured relative to the channel width $W$) is selected, just as for molecular fluids\cite{PhysRevLett.56.2036}. In our experimental setup, as the finger advances in the channel, its tip velocity $U$ slowly increases and the relative finger width $\lambda$ decreases. \Figc{figxyfinger} shows results for $\lambda$ as a function of the control parameter $1/B$. Different experiments for different aspect ratios $\epsilon$ collapse. The finger width goes below the 1/2-limit, which is also seen in experiments with molecular fluids \cite{Tabeling_Zocchi_Libchaber_1987} and recent simulations \cite{ledesma_aguilar_pagonabarraga_2} albeit at much higher control parameters. Visco-elastic effects cannot be completely ruled out: shear-thinning fluids typically display finger narrowing, see for example \cite{lindner09} and references therein. However, as noted above, typical shear rates in our microfluidics experiments lie well below the onset of shear thinning as determined rheologically. For comparison, we also plot the theoretical curve by McLean and Saffman \cite{mclean81}, which tends to 1/2 at large $1/B$. Finally, we observe that the finger shape is well described by the semi-empirical Pitts equation\cite{Pitts_1980}, which reads $\cos (\pi y/ 2 \lambda) = \exp (\pi x/2 \lambda)$, despite the complex nature of our fluids, see \figb{figxyfinger}. 

\begin{figure*}
\begin{centering}
\includegraphics[width=11cm]{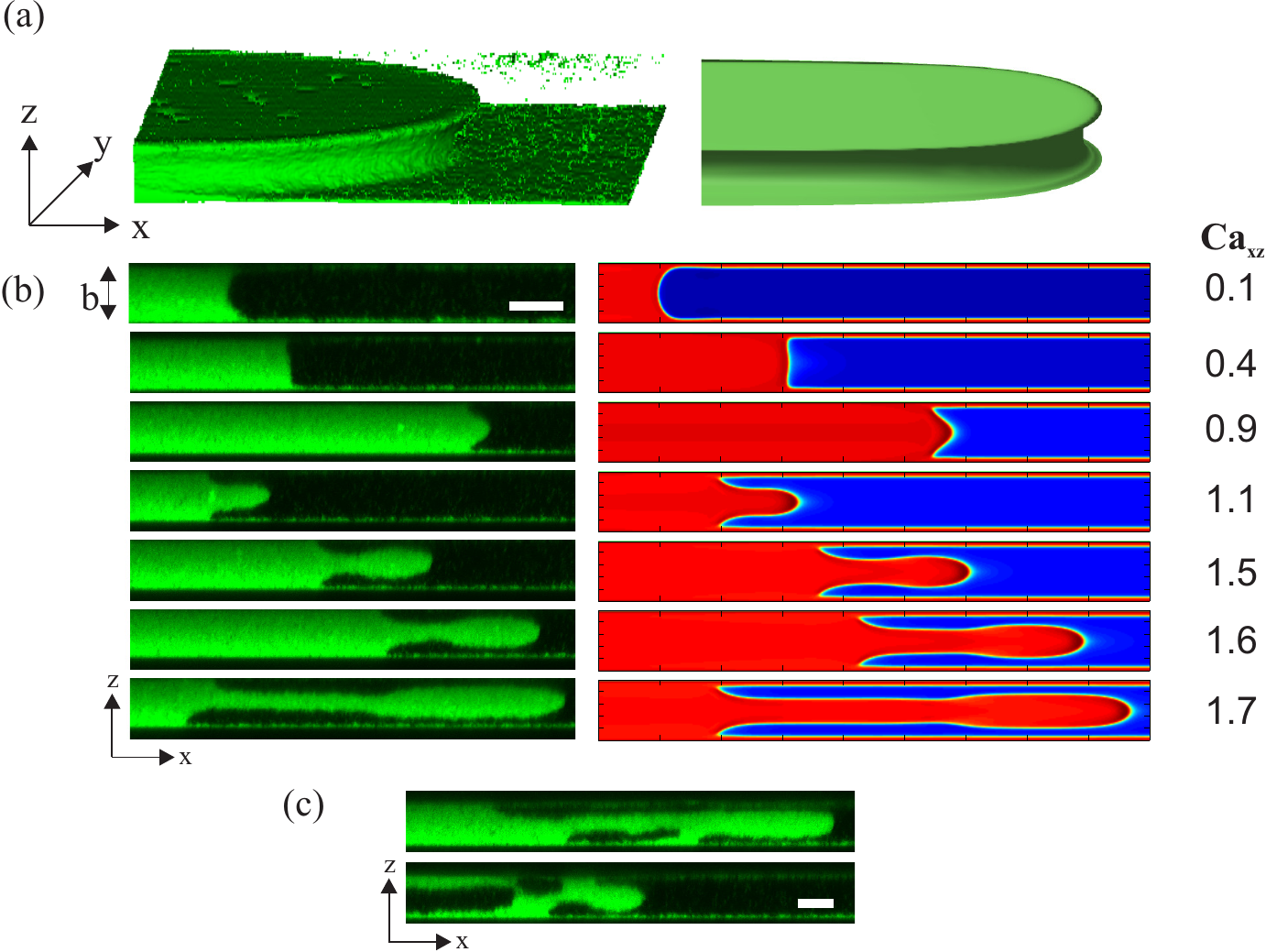}
\caption{(a) $3D$ images of a single finger from experiments (left panel)  and lattice Boltzmann simulations (right panel). (b) Profile of the finger in the $xz$-plane as a function of the capillary number $Ca_{xz}$.  LSCM images (left) of the leading interface are compared to the simulations (right). The loss in fluorescence intensity as a function of $z$ is due to the refractive index mismatch between colloids and solvent. (c) LSCM images of fingers merging with the top or bottom wetting layers at large capillary numbers, before being pulled back to the centre of the channel. Scale bars indicate 10 $\mu$m.}
\label{comparison}
\end{centering}
\end{figure*}

{\bf (b) Observations in the $xz$-plane:}
As explained in the introduction, discrepancies between the measured finger width for molecular fluids and the theoretical prediction by McLean and Saffman, were explained by taking the curvature of the advancing meniscus in the $xz$-plane into account \cite{Tabeling_Libchaber1986}. Although this will not rescale our measured finger widths above the 1/2-limit --only the $x$-axis is rescaled-- it is interesting to study the shape of the interface in the $xz$-plane. It turns out that the behaviour of the meniscus is surprisingly rich. 

\Figa{comparison} illustrates that our combination of system and techniques allows observing the Saffman-Taylor fingers fully in 3D. The left column of \figb{comparison} shows the $xz$-interfacial profile at different capillary numbers. The data are obtained by making an $xz$-cut through the tip of the advancing finger. At low velocities the interface advances as a concave meniscus. This is a consequence of the wetting properties of the colloid-polymer mixtures: due to the depletion interaction the colloids are strongly attracted to hard, smooth walls \cite{aarts05jpcb,jamiejcp12} and this leads to the colloidal-rich `liquid' phase wetting the microfluidic walls completely. At a capillary number $Ca_{xz} = (\eta_1 + \eta_2 ) U / 2 \gamma \sim 0.4$, the $xz$-interfacial profile in the centre of the channel  is flat, i.e. its curvature is zero. Note that a scaling analysis \cite{thesisyiannis} shows that the relevant capillary number now depends on the average viscosity, not the viscosity difference as in $Ca_{xy}$. Then, at higher $Ca_{xz}$ the meniscus first becomes concave, and at $Ca_{xy} \sim 1$, the interface position becomes decoupled from that of the contact line\footnote{As we are in the complete wetting regime, the liquid slides over a colloid-rich wetting layer such that there is no true contact line. However, there is an obvious line where the curved meniscus meets the surface layer, for which we use the term contact line.}, and a finger in the $xz$-plane is formed, which we will call an $xz$-finger. We stress that this is due to a completely different mechanism than the viscous fingering in the $xy$-plane; it is not driven by an instability of the interface, but rather by the inability of the contact line to move as quickly as the leading interface. The $xz$-finger is of a characteristic shape, with a narrowing near the contact line. 
At yet higher capillary numbers a meandering motion of the finger is observed, where it makes contact with either the top or bottom wall (\figc{comparison}), before continuing to move freely again in the centre of the channel. This is a consequence of the large thermal interface fluctuations in colloid-polymer mixtures and therefore intimately related to the ultralow surface tension of our system. The advancing $xz$-finger can `feel' the wetting layers and thus bridge the gap until the mismatch in tip vs.\ contact line velocity brings the $xz$-finger into the center of the channel again. The meandering entraps bubbles of the high viscosity phase in the advancing finger.


To further interpret the experiments we used a free energy lattice Boltzmann algorithm\cite{PhysRevLett.75.830,PhysRevE.54.5041} to solve the hydrodynamic equations of motion for a binary fluid (Model H). In order to study the penetration process in the $xz$-plane we considered a channel formed by 2 plates of infinite width, parallel to the $xy$-plane, located at $z=0$ and $z=b$. Channel dimensions were, in simulation units, $b=50-100$ and length $L=1000-2000$, whereas the thickness of the wetting layer was $\zeta \sim 5$. A constant body force was applied to both phases. We matched the capillary and Peclet numbers and the viscosity ratio to the experimental situation. 
The second column of \figb{comparison} presents results from lattice Boltzmann (LB) simulations showing how the $xz$-interfacial profile responds to the flow, in close correspondence to the experiments at the same capillary number. When using the LB approach to model interfaces in molecular fluids, discretization constraints mean that the diffuse interface is unphysically wide and care must be taken that this does not affect the results \cite{PhysRevE.69.031603}. Here, however, the relatively large interfacial width of our colloid-polymer mixtures \cite{hennequinprl} allows a quantitative match between experiment and simulation: excellent agreement is found. 
In the simulations the hydrodynamic singularity at the contact line is alleviated by interdiffusion, driven by the chemical potential imbalance as the interface is pulled out of thermodynamic equilibrium. In the experiments it is likely (although not proven) that the mechanism for contact line motion is the same.

\begin{figure}
\includegraphics[width=75mm]{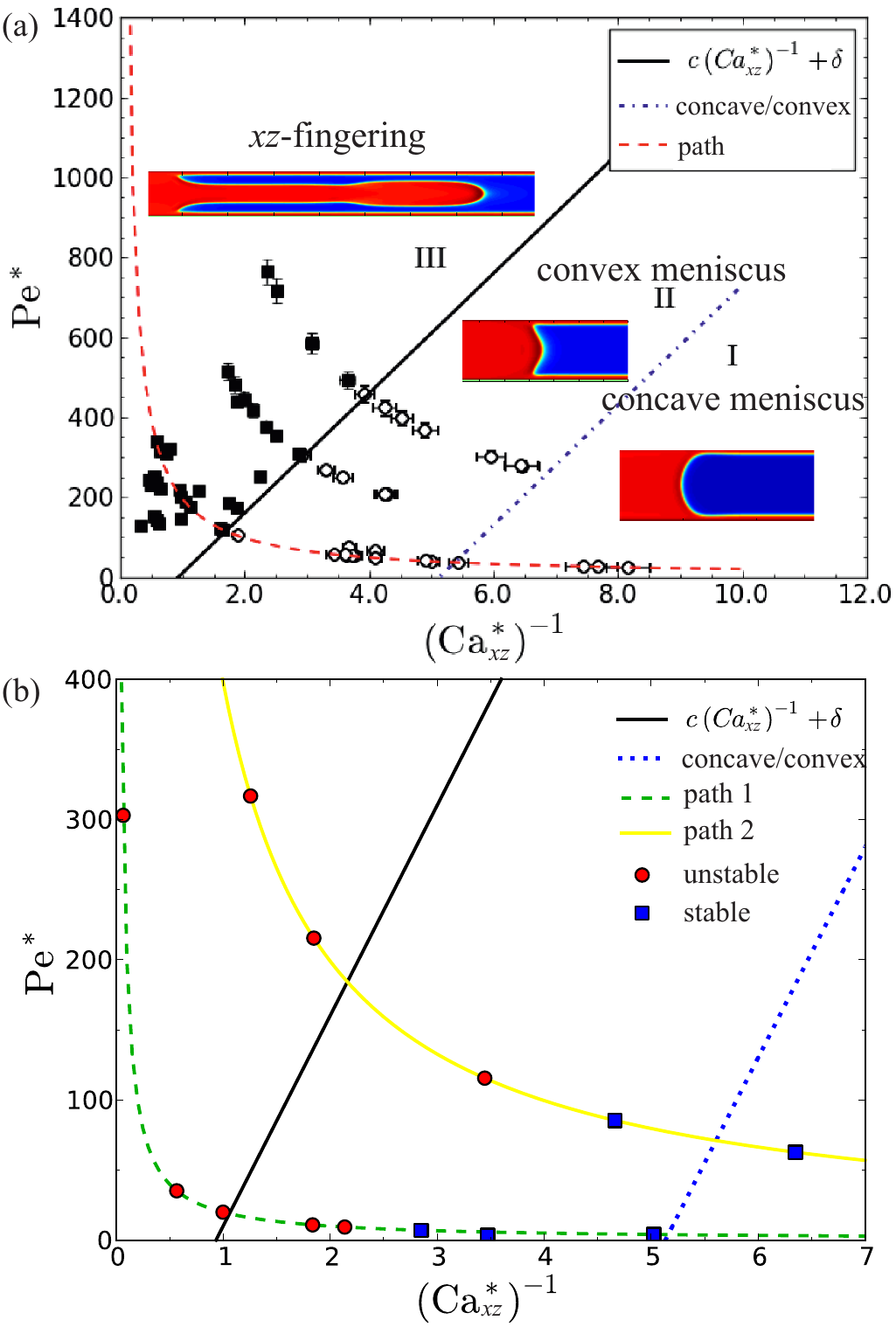}
\caption{(a) The shape of the interface in the $xz$-plane for a viscosity contrast $r=0.4$ as a function of the scaled capillary and Peclet numbers, $Ca^*_{xz}$ and $Pe^*$, all matched to the experiments. The snapshots represent the different configurations ($\circ$ meniscus regime, $\blacksquare$ $xz$-fingering regime). The solid line corresponds to the onset of fingering in the $xz$-plane. The dash-dotted line indicates a flat advancing interface. The red/dashed line describes a possible experimental path, where $Pe^{*}$ and $Ca^*_{xz}$ are tuned by varying only the leading interface velocity $U$. 
(b) Simulation results for two different, possible experimental paths in the $(Pe^*,(Ca^{*}_{xz})^{-1})$-plane with $Pe^*/Ca^*_{xz} = 19.9$ (path 1) and $398.0$ (path 2) for $r = 0.4$. Red circles denote the results where perturbations grow and lead to viscous fingers in the $xy$-plane (exponential growth of the amplitude of the perturbations), and blue squares denote results with a negative growth rate. The solid black line corresponds to the onset of fingering in the $xz$-plane, while the blue/dash-dotted line indicates a flat advancing interface.
}
\label{simulations}
\end{figure}

Ledesma {\em et al.}\cite{ledesma_aguilar_pagonabarraga_1,ledesma_aguilar_pagonabarraga_2} have argued, based on a scaling argument relating the velocity of the leading interface to that of the contact line, that for fluids at neutral wetting (i.e.\ a contact angle of 90$^\circ$), fingering in the $xz$-plane occurs for $Pe > c(Ca_{xz})^{-1}$, where the Peclet number $Pe=Ub/2D$, with $D$ is the diffusivity of the concentration field, and $c$ is a constant of order 1 . An extension of their argument to complete wetting and fluids of different viscosity gives \cite{thesisyiannis}
\begin{equation}
\label{eqvf}
Pe^* > c (Ca_{xz}^* )^{-1}+\delta, 
\end{equation}
where the rescaled Peclet and capillary numbers, $Pe^{*} = U (b/2-\zeta)/D$ and $Ca_{xz}^{*}=\{1-4 (\zeta /b) (1-\zeta/b)\}Ca_{xz}$, and the offset $\delta$, account for the thinkness $\zeta$ of the wetting layer at the surfaces. A full theoretical derivation will be given elsewhere \cite{yiannisfuture}. 

This implies that the dynamics can be captured in the $(Pe^*,(Ca_{xz}^*)^{-1})$ plane (\fig{simulations}). \Figa{simulations} shows numerical results confirming the validity of the formula above for a viscosity ratio $r=0.4$ with $c = 150$ and $\delta = -140$, denoted by the full line. To the right of the line the system moves in the meniscus regime, to the left of the line $xz$-fingers develop. Note that $c$ takes a value two orders of magnitude larger than that for neutrally wetting molecular fluids. This is entirely due to the wetting properties of our fluids.  The increased ease with which the contact line can move across the pre-formed wetting layer means that fingering in the $xz$-plane is strongly suppressed. 

In our experiments the velocity $U$ is slowly increasing during each experimental run, which means that we scan the $(Pe^*,(Ca_{xz}^*)^{-1})$-phase diagram by  moving along the dashed red line in \figa{simulations} from the bottom right corner (low velocity) to the top left corner (high velocity). Note that $Pe^*$, $Ca^*_{xz}$  and $r$ have been matched to the experiments. The interface thus first moves with a concave meniscus, at intermediate velocities it moves with a convex meniscus and finally, and after crossing the full straight lines in \figa{simulations} it displays $xz$-fingering. In our experiments we find that fingering in the $xz$-plane commences at $1/Ca^*_{xz} \sim 1.6$ and $Pe^* \sim 110$, where we have used the diffusion coefficient of the colloids to estimate the Peclet number. This compares favourably to $Pe^* \sim 100$ at $1/Ca^*_{xz} \sim 1.6$ found in the simulations.

Simulations further reveal that the onset of fingering in the $xz$-plane varies with viscosity ratio $r$. As the difference in viscosity of the two fluids increases, we find that fingering in the $xz$-plane is even more strongly suppressed: much higher capillary numbers are needed both to change the shape of the meniscus from concave to convex and to induce the formation of a finger. Numerically, we find $c = 350$, $\delta = -75$ for $r=0.90$, while $c = 28$ and $\delta = -38$ for $r=0$ \cite{thesisyiannis}.


{\bf (c) Interplay between interfacial curvature in the $xz$-plane and $xy$-fingering:}
The above has important consequences for the viscous fingering instability itself.  
In particular, the simulations reveal that viscous fingering in the $xy$-plane is suppressed when the meniscus in the $xz$-plane is concave. This is evident from \figb{simulations}, where we plot results based on 3D simulations for two systems with different values for $Pe^*/Ca^*_{xz}$ at a given viscosity contrast and aspect ratio. Red circles denote points where an interface that has been subjected to a small perturbation in the $xy$-plane led to viscous fingers, i.e.\ a positive growth rate such that the amplitude of the perturbation grows exponentially rendering the interface unstable, while blue squares denote the points where the perturbation died out and, therefore, did not develop any fingering (negative growth rate - stable interface). 

This demonstrates that the onset of viscous fingering in the $xy$-plane indeed depends strongly on the behaviour of the leading interface in the $xz$-plane. The concavity of the interface in the $xz$-plane suppresses the onset of viscous fingering in the $xy$-plane.  Hence, it is unlikely to be possible to scale the finger width to the McLean-Saffman solutions by only introducing a modification of the control parameter $1/B$. Furthermore, full dispersion relations, which will be published elsewhere \cite{yiannisfuture}, now depend on the scaled capillary and P{\'e}clet numbers to take the curvature in the $xz$-plane into account, and on the channel's aspect ratio.
All this underlines the importance of the third dimension in the viscous fingering instability.

\section{Conclusion}
\label{concl}
In summary, we have studied the development of viscous fingers in colloid-polymer mixtures, both in experiment and in simulations. We set out to discuss these experiments in light of three topics: (i) Despite the ultralow surface tension between the two demixed phases, the fingering displays the classic phenomena. A single finger width is selected as a function of $1/B$ and the finger contour is well described by the Pitts equation. Moreover, LB simulations, which satisfy continuum hydrodynamics, describe the instability semi-quantitatively in terms of dimensionless fluid parameters. Only at larger driving velocities do we see a meandering motion of the $xz$-fingers in the experiments, but not in the simulations, which is a consequence of the thermal interface noise. (ii) The combination of LSCM and microfluidics has enabled us to observe the fingering instability in three dimensions. We contrast fingering in the $xz$-plane, caused by the inability of the contact line to move at sufficient velocity to keep up with the leading interface, and fingering in the $xy$-plane, due to the interface instability identified by Saffman and Taylor. 
(iii) Our displacing fluid wets the walls of the microfluidic device. As a result the curvature of the meniscus in the $xz$-plane is a function of the driving velocity. LB simulations subsequently reveal that the cross-over from a concave to a convex meniscus, and to $xz$-fingers, depends on the capillary and Peclet numbers, and on the viscosity contrast. The interfacial curvature in the $xz$-plane has a pronounced effect on the onset of the Saffman-Taylor instability. In particular, there is a threshold velocity for viscous fingering. 

The agreement so far between experiments and simulations is excellent. However, a detailed experimental verification of the predicted suppression of the instability is still needed, although preliminary experiments for other statepoints strongly support the predictions. 

Finally, we note that our results may be exploited in enhanced oil recovery\cite{DangVu200980}, where the wettability of the displacing fluid (water and surfactants) can be tuned to displace the oil most effectively. 

{\bf Acknowledgments} We thank Rodrigo Ledesma-Aguilar, Daniel Bonn and Roel Dullens for useful discussions. We acknowledge financial support from Research Promotion Foundation RPF (Republic of Cyprus) under project PENEK/SUPPORT/0308/17 (Co-financed by the European Regional Development Fund), from the Ministry of Higher Education Malaysia (MOHE), from EPSRC (EP/H035362) and from the ERC advanced grant MICE.

\providecommand*{\mcitethebibliography}{\thebibliography}
\csname @ifundefined\endcsname{endmcitethebibliography}
{\let\endmcitethebibliography\endthebibliography}{}

\end{document}